\documentclass{aa}

\usepackage{soul}
\usepackage[T1]{fontenc}
\usepackage{txfonts}
\usepackage{graphicx}	
\usepackage{amsmath}	
\usepackage{amssymb}	
\usepackage{bm}
\usepackage[hidelinks,colorlinks=true,linkcolor=blue,citecolor=blue]{hyperref}
\usepackage[normalem]{ulem}

\newcommand{\be}{\begin{equation}}
\newcommand{\ee}{\end{equation}}

\begin{document} 

\title{Polarization properties of synchrotron sources from simulations of relativistic magnetohydrodynamic turbulence}

\authorrunning{L. Del Zanna et al.}
\titlerunning{Polarization properties of synchrotron sources from simulations of RMHD turbulence}

\author{
Luca Del Zanna \inst{1,2,3}
\fnmsep\thanks{luca.delzanna@unifi.it}
\and 
Niccolò Bucciantini \inst{2,1,3}
\and 
Simone Landi \inst{1,2,3}
}

\institute{
Dipartimento di Fisica e Astronomia, Universit\`a di Firenze, Largo E. Fermi 2, I-50125 Firenze, Italy
\and
INAF, Osservatorio Astrofisico di Arcetri, Largo E. Fermi 5, I-50125 Firenze, Italy
\and
INFN , Sezione di Firenze, Via G. Sansone 1, I-50019 Sesto Fiorentino (FI), Italy
}

\date{Received 4 July 2025; accepted 5 September 2025}

 
  \abstract
   {}
   {The emission from the relativistically hot plasmas of high-energy astrophysical synchrotron sources, pulsar wind nebulae (PWNe) in particular, depends on the level of magnetic fluctuations. Recent observations by the X-ray polarimeter IXPE support the presence of turbulence, with varying conditions even in different regions of a same source. We model such emission, and in particular the degree of linear polarization, by using 3D relativistic magnetohydrodynamic (MHD) turbulence simulations for the first time.}
   {Thanks to a novel accelerated version of the \texttt{ECHO} code, a series of 3D relativistic MHD simulations were performed assuming a relativistically hot plasma and various degrees of magnetization, mimicking different conditions encountered in synchrotron sources. Magnetic fluctuations at random directions with respect to a background field were initialized at large scales. After the full development of the turbulent cascade, the statistical properties of the plasma and of the synchrotron emission maps were analyzed.}
   {Turbulence rapidly relaxes to a sort of Alfvénic equilibrium and a Kolmogorov cascade with a slope of $-5/3$ soon develops, with differences depending on the initial ratio, $\eta$, of magnetic fluctuations over the background field. Dissipation mostly occurs in thin current sheets, where (numerical) reconnection takes place and intermittency and deviation from isotropic Gaussian distributions are observed. Synthetic synchrotron maps and their statistical properties depend on $\eta$ too, approaching analytical estimates for large $\eta$. The integrated degree of linear polarization is found to cover the whole range of observed values in PWNe, and its dependence on the relative amplitude of turbulent fluctuations shows a good agreement with analytical estimates, even in the presence of anisotropy.}
   {}


   \maketitle

\section{Introduction}

High-energy astrophysical sources of synchrotron emission are characterized by the presence of ultra-relativistic electrons (pairs in general) in a magnetized plasma. Examples range from the jets and radio lobes of active galactic nuclei (AGNs) \citep[since][]{Baade:1956,Burbidge1956} to those of microquasars \citep{Mirabel1999}, from the regions surrounding supermassive black holes \citep[responsible for the famous images of M87$^*$ and Sgr A$^*$,][]{EHT2021,EHT2024} to gamma-ray bursts (GRBs) \citep{Lloyd2000,Burgess2020}, and from supernova remnants (SNRs) \citep{Reynolds2008} to the radio emission by cosmic rays in the Galactic interstellar magnetic field \citep{Orlando2013,Planck2016}. 

However, the most representative class of such astrophysical sources is definitely provided by pulsar wind nebulae (PWNe) \citep{Rees1974,Kennel1984a,Kennel1984b,Gaensler2006,Hester2008,Buehler2014}, bubbles of relativistically hot plasma confined by an external medium, either the remnant of the parent pre-supernova progenitor, or, for older systems in the so-called bow-shock phase, the interstellar medium itself. The PWNe are energized by the wind from a rapidly rotating and strongly magnetized neutron stars (a pulsar), the relic of the stellar explosion. Such objects, in particular the Crab nebula, are prototypical examples of optically thin synchrotron-emitting sources, from radio to mega-electronvolt gamma-rays. They are also characterized by inverse Compton emission (by diffuse dust emission, stellar radiation field, cosmic microwave background or even synchrotron light itself, as in the Crab nebula) in the giga- to tera-electronvolt bands and beyond. It was indeed thanks to the high level of optical polarization measured in the Crab nebula \citep{Oort1956,Woltjer1957} that synchrotron emission was recognized for the first time as an astrophysical radiation process, as was previously suggested by \cite{Shklovskii1953}.

The large-scale structure of the magnetic field inside a synchrotron source is obviously very important, shaping the overall appearance of the object, and X-rays are able to better investigate regions where particles are accelerated, due to shorter cooling times. For example, X-ray observations of the inner part of the Crab nebula by the Chandra satellite revealed for the first time a dominant ``jet-torus'' structure and additional finer details \citep{Weisskopf2000}, later recognized also in the optical band through observations by the Hubble Space Telescope (HST) \citep{Hester2002}. Axisymmetric relativistic magnetohydrodynamic (MHD) simulations managed to reproduce these observations, assuming a mainly toroidal magnetic field and a stronger pulsar wind energy flux toward the equator, resulting in an oblate shape of the wind termination shock \citep{Komissarov2003,Komissarov2004,DelZanna2004}. Assuming power-law distributions of relativistic particles, either accelerated at the termination shock \citep{Sironi2011} and/or through reconnection inside the nebula \citep{Sironi2014}, synthetic nonthermal emission maps and spectra were also computed on top of these models, from radio to gamma rays, reproducing fine structures such as rings, moving ``wisps'', and the ``knot'' \citep{DelZanna2006,Volpi2008,Camus2009,Olmi2014,Olmi2015}.

The picture changes in 3D simulations, in which a magnetization parameter close to unity is allowed in the pulsar wind, the nebular magnetic field is no longer purely azimuthal, poloidal components are present especially along the polar jets, subject to kinks, and strong mixing and dissipation occur \citep{Porth2014,Olmi2016}. Although there is a clear indication that the field has a more complex structure, the limited numerical resolution prevents one from accurately modeling what happens at smaller scales, where turbulence is expected to develop and play a role, especially for particle transport and diffusion \citep{Tang2012,Porth2016}.

Precise information on the large-scale structure of the nebular magnetic field can only be provided by polarimetric observations, recalling that synchrotron emission is known to be mainly linearly polarized with respect to the direction of the field (projected on the plane of the sky). The maximum expected polarization degree (PD) is \citep{Ginzburg1965}
\be
\label{eq:Pi_max}
\Pi_{\rm max} = \frac{p+1}{p+7/3} = \frac{\alpha+1}{\alpha+5/3} \simeq 70\%,
\ee
a fraction holding for standard values of the slope of the electron energy distribution function, $N(\epsilon) \mathrm{d}\epsilon\propto \epsilon^{-p}\mathrm{d}\epsilon$, typically with $p=2.0 - 2.2$, corresponding to an emissivity $j_\nu\propto \nu^{-\alpha}$ with $\alpha=(p-1)/2=0.5 - 0.6$ \citep{Veron-Cetty1993}, and assuming pitch angle isotropy. 

This limit is not easily reached, however, as in any realistic source small-scale turbulent magnetic fluctuations are expected to depolarize the emission. The dependence of radio synchrotron emission, in particular its polarization, on the MHD turbulence properties has been vastly investigated in the case of the Galactic interstellar medium, and various techniques have been proposed to infer these properties from observations \citep[e.g.][]{Lazarian2012,Lazarian2016,Lazarian2024}.
On top of this, in sources characterized by a structured magnetic geometry, there are also depolarization effects due to line-of-sight and/or spatial integration \citep{Bucciantini2017b}. 

At present the sources displaying the highest level of polarized synchrotron emission are SNRs and PWNe. These objects, moreover, are typically well resolved, especially in the radio band, allowing us to perform quite detailed diagnostics of the local magnetic field geometry \citep{Dubner2015}. An interesting dichotomy appears to characterize SNRs: young sources show a magnetic field that is mainly radial, probably due to the presence of instabilities, whereas in more evolved ones the field is approximately tangential to the shock surface, as is expected from pure compression. In the case of PWNe, radio observations maps of the Crab nebula show a rather low PD, of about $10\%$, and it is even worse where interaction with the external filaments protruding from the slowly expanding remnant takes place \citep{Wilson1972,Velusamy1985}, while optical measurements have always shown higher values of about $20\%$ and above closer to the center \citep{Schmidt1979}, confirmed by the first X-ray measurement \citep{Weisskopf1978}. Values of the PDs close to the maximum theoretical limit, $\Pi_{\rm max}$, were reported using HST data in the finest structures such as the knot and wisps \citep{Moran2013}, suggesting that the inner regions and features possess a much more uniform magnetic field.

In order to spatially resolve the magnetic structure of high-energy astrophysical sources, in 2021 the ``Imaging X-ray Polarimetry Explorer'' (IXPE) satellite was launched, and results for the Crab nebula confirm an inner predominant toroidal field, an integrated PD of $\Pi \sim 20\%$ in the nebula, showing, however, a rather patchy distribution, with strong variations and peaks of $45-50\%$ \citep{Bucciantini2023,Wong2023}. Even higher values, of $\Pi\sim 60\%$, are found in the inner region of Vela PWN \citep{Liu2023,Deng2024}, which is a more evolved source in the so-called reverberation phase of interaction with the reverse shock of the remnant. Different magnetic structures are found by IXPE in the synchrotron-emitting shells of SNRs, where the direction is mainly radial, and again various PD values are measured indicating different levels of depolarization by small-scale turbulence \citep{Vink2022,Ferrazzoli2023}. For a recent summary and discussion on IXPE results in various sources, see \cite{Bucciantini2024}.

As far as models are concerned, if one wants to reproduce such a variety of PDs in different objects or in different regions of a same source, turbulence must be invoked. In the case of PWNe, this is probably induced by the motions around the termination shock, causing the variable wisps \citep{Camus2009}, or by the kinks of the magnetized jets as has been shown in 3D simulations \citep{Porth2014,Olmi2016}, or even by the interaction with the external ambient, through the nonlinear development of (magnetic) Rayleigh-Taylor and/or Kelvin-Helhmoltz instabilities \citep{Bucciantini2004,Bucciantini2006}. By applying analytical corrections to the expected synchrotron emission in the presence of magnetic fluctuations, recipes for computing the Stokes parameters, emissivity, and PD have been provided in the case of SNRs \citep{Burn1966,Bandiera2016,Bandiera2024}, and applied to an analytical toy model of PWNe, characterized by a torus of emitting plasma with radial flow and a toroidal magnetic field \citep{Bucciantini2017}. The comparison of synthetic results with observations by Chandra suggests the presence of a significant turbulent component of the magnetic field, whose energy could be at least $1.5$ times larger than the ordered one to explain depolarization \citep[see also][]{Nakamura2007,Mizuno2023}.

Unfortunately, the presence of small-scale turbulence is difficult to account for in global numerical simulations, due to limited resolution; hence, even axisymmetric models of the polarized emission of PWNe have necessarily been based so far on a rather laminar field \citep{Bucciantini2005,DelZanna2006}. In these works, a detailed model of the synchrotron emission was employed, assuming a power law of emitting electrons with a given index and number density proportional to the local energy density (dominated by  thermal pressure). The polarized emission was computed by assuming an optically thin plasma and by taking into account the local direction of the magnetic field with respect to the line of sight (LOS), the transformation of vector quantities from the comoving frame of the fluid to the observer's frame, and in particular the relativistic Doppler boosting and polarization angle swing effects, which may complicate the diagnostics in the case of fast flows, especially near the termination shock.

In the present work, we apply the same general recipes to compute the synchrotron emission and polarization properties, for the first time on top of data coming from numerical simulations of decaying turbulence in a relativistically hot, magnetized plasma, in order to investigate the effects on the polarized emission of magnetic fluctuations directly induced by turbulence. Our goal is to analyze, on the one hand the properties of relativistic turbulence in such an environment, and on the other hand the statistical behavior of synthetic synchtrotron emission maps, distributions of Stokes parameters, and the PD. Moreover, we check whether the crude approximations of perfectly isotropic and Gaussian distributions for magnetic stochastic fluctuations, which were previously assumed in the cited analytical models in the case of SNRs, are far from our turbulence results, at all scales, and whether the inferred PDs are close to the analytical ones or not.

To achieve these goals, we perform, using a novel GPU-accelerated version of the \texttt{ECHO} code for (general) relativistic MHD \citep{DelZanna2007,DelZanna2024}, a series of 3D simulations of decaying turbulence. We actually consider a localized fraction of the emitting torus of a PWN, rather than the global source, simply modeled as a cubic, periodic numerical box. The turbulent cascade of fluctuations in the plasma is followed from the large injection scales down to the (numerical) dissipative ones, assuming a relativistic MHD regime and hot plasma conditions, as appropriate for PWNe. This is done for several initial amplitudes of the magnetic fluctuations, assumed to have random directions, and of the uniform background field, as well as assuming a LOS parallel or perpendicular to it. 

In the present paper, Section 2 is devoted to numerical simulations and analysis of turbulence, and Section 3 to the analysis of results of synthetic synchrotron polarized emission. Conclusions are drawn in Section 4.

\section{Relativistic 3D MHD turbulence simulations}

The ideal, special relativistic MHD equations were solved using the novel GPU-accelerated version of the \texttt{ECHO} code \citep{DelZanna2007}. We refer to \cite{DelZanna2024} for both the description of the system of equations, and for the acceleration strategies, based on the simple exploitation of the ``Standard Language Parallelism'' capabilities of modern \texttt{FORTRAN} and the ``Unified Memory'' paradigm of the NVIDIA compiler. Performances such as the speedup of GPUs over CPUs (up to $\times\, 38$, according to our latest estimates), and strong and weak scaling, are reported too. These were tested on the Leonardo supercomputer at CINECA (Bologna, Italy), where the simulations described here were also run.

In the remainder of the paper we assume a Minkowskian flat spacetime and Cartesian coordinates. The numerical domain is a periodic, cubic box with size $(2\pi)^3$, where all lengths (and the coordinates $\bm{x}$) are thus normalized against a generic spatial dimension, $L=1$. We also let $c=1$ and absorb the factor $1/\sqrt{4\pi}$ in the definition of the electromagnetic field, as is usually assumed in relativistic MHD codes.

\subsection{Setup of fluctuations and choice of parameters}

In the present subsection we describe the setup for our simulations of 3D-MHD decaying - that is without an external driver - turbulence in relativistically hot and strongly magnetized plasmas, and we define the parameters characterizing the various runs. The plasma is assumed to be initially static, in all cases.

Relativistically hot plasmas are characterized by the condition $p\gg \rho$; that is, the thermal pressure is dominant with respect to the rest mass energy density. The inertia is due to the relativistic enthalpy, the sum of total energy density and pressure, $\rho h = \rho + \epsilon +p$ ($h$ is the specific enthalpy and $\epsilon$ as the internal thermal energy density). For an ideal gas we have $p = (\gamma - 1)\epsilon$, and for an adiabatic index of  $\gamma=4/3$, as is appropriate for a relativistic gas, we find $h = 1 + 4p/\rho$. When $p\gg \rho$ we find $\rho h \to 4p$, and this also leads to the highest possible value for the relativistic sound speed: $c_s = \sqrt{\gamma p / \rho h} \to 1/\sqrt{3}$. 

As far as the magnetic field is concerned, we assume a background uniform field, $\bm{B}_0$, and random 3D fluctuations both along and perpendicular to it, of a given rms (root mean square) strength, $B_1$. At the initial time, $t=0$, we set up the fluctuations imposing only long wavelength Fourier modes with wave vector components satisfying $-4\leq k_i \leq 4$, in all directions, $i=1,2,3$; that is
\be
\delta\bm{B} = B_1 \sum_{\bm{k}=-4}^{4} 
\cos (\bm{k}\cdot\bm{x} + \varphi) (\cos\vartheta \bm{e}_1 + \sin\vartheta \bm{e}_2),
\label{eq:setup}
\ee
where we introduced two new unit vectors as
\be
\bm{e}_1 = \frac{\bm{k}\times\bm{B}_0}{|\bm{k}\times\bm{B}_0|}, \qquad
\bm{e}_2 = \frac{\bm{k}\times\bm{e}_1}{|\bm{k}|},
\ee
normal to each other and both normal to the wave vector, $\bf{k}$, to preserve automatically the solenoidal condition, $\nabla\cdot\delta {\bf B} = 0$. The phase, $\varphi$, and the position angle, $\vartheta$, in the $\bm{e}_1 - \bm{e}_2$ plane were chosen randomly for each $\bf{k}$. The magnetic fluctuations computed above do not necessarily lie in the plane orthogonal to $\bm{B}_0$, unless one forces $\vartheta =0$. Notice that the constant $B_1$ can actually only be computed after all random components have been introduced, as we want to impose
\be
B_1 = B_\mathrm{rms} \equiv \sqrt{< (\delta \bm{B} )^2>}\,,
\ee
where the spatial average spans the whole numerical domain. In the remainder we actually derive the ordered and disordered initial magnetic field constants, $B_0$ and $B_1$, from the definitions
\be
\sigma_0 = \frac{B_0^2}{\rho h} , \qquad 
\sigma_1 = \frac{B_1^2}{\rho h} ,
\ee
which are the respective magnetization parameters. Notice that we are using the so-called hot definition, appropriate when $p\gg \rho$, whereas the cold definition assumes $h=1$ \citep[e.g.][]{DelZanna2016}. Other important plasma parameters are the relativistic Alfv\'en speed and the ``plasma beta'', defined, respectively, as
\be
c_a = \frac{B}{\sqrt{\mathcal{E}}}, \qquad \beta = \frac{2p}{B^2},
\ee
where $\mathcal{E}=\rho h + B^2$ takes into account all forms of inertia. Recalling that in our case $\rho h \to 4p$, the total plasma magnetization, $B^2/\rho h$, is simply proportional to $\beta^{-1}$. Moreover, $c_a$ can clearly approach unity (that is $c$) in regimes of strong magnetization, and given that at $t=0$ we have
\be
B = \sqrt{B_0^2 + B_1^2} = \sqrt{\rho h (\sigma_0 + \sigma_1)} \,,
\ee
this is expected to occur when either $\sigma_0\gg 1$ or $\sigma_1\gg 1$.

\begin{table}
\caption{Magnetic field parameters for all turbulence simulations, listed for increasing values of the initial relative amplitude $\eta = B_1/B_0$.}
\label{Tab:param}
{\renewcommand{\arraystretch}{1.3}
\begin{tabular}{c c c c c c c c} 
\hline
 & $\sigma_0$ &  $\sigma_1$ & $B_0^2$ & $B_1^2$ & $\eta$ & $c_a$ & $\beta$ \\  
\hline
\texttt{R01}  & $1$     & $0.05$ & $1000$ & $50$ & $ 0.224$ & $0.716$ & $0.476$ \\
\texttt{R02}  & $1$     & $0.1$ & $1000$ & $100$ & $ 0.316$ & $0.724$ & $0.455$ \\
\texttt{R03}  & $1$     & $0.2$ & $1000$ & $200$ & $0.447$  & $0.739$ & $0.417$ \\
\texttt{R04}  & $0.1$   & $0.05$ & $100$ & $50$ & $0.707$  & $0.361$ & $3.33$ \\
\texttt{R05}  & $0.1$   & $0.1$ & $100$  & $100$ & $1.00$   & $0.408$ & $2.50$ \\
\texttt{R06}  & $0.1$   & $0.2$ & $100$  & $200$ & $1.41$   & $0.480$ & $1.67$  \\
\texttt{R07}  & $0.01$  & $0.05$ & $10$  & $50$ & $2.24$   & $0.238$ & $8.33$  \\
\texttt{R08}  & $0.01$  & $0.1$ & $10$   & $100$ & $3.16$   & $0.315$ & $4.55$ \\
\texttt{R09}  & $0.01$  & $0.2$ & $10$   & $200$ & $4.47$   & $0.417$ & $2.38$ \\
\texttt{R10} & $0.001$ & $0.05$ & $1$   & $50$ & $7.07$   & $0.220$ & $9.80$ \\
\texttt{R11} & $0.001$ & $0.1$ & $1$    & $100$ & $10.0$   & $0.303$ & $4.95$ \\
\texttt{R12} & $0.001$ & $0.2$ & $1$    & $200$ & $14.1$   & $0.409$ & $2.49$ \\
\hline
\end{tabular}}
\end{table}

In all simulations of decaying turbulence, we normalized quantities against a background constant rest mass density $\rho = 1$ and assumed a relativistically hot gas with $h=1000$ and $p\simeq 250$ (that is $p=\rho h/4$), for a constant sound speed, $c_s\simeq 0.577$ (the highest possible value). The magnetic parameters were varied according to table \ref{Tab:param}, listed for increasing values of the normalized amplitude of fluctuations
\be
\eta=\frac{B_1}{B_0} = \sqrt{\frac{\sigma_1}{\sigma_0}},
\ee
which can be either smaller or larger than unity. Notice that we fixed the fluctuations to be quite high, from $B_1^2=50$ to $200$, while the background field decreases from $B_0^2=1000$ down to unity (values equivalent to the cold magnetization parameter, while the hot magnetization, $\sigma_0$, ranges from $1$ to $0.001$).

Simulations followed the evolution of the decaying turbulence and the rms values of characteristic quantities such as the curl of magnetic and velocity fluctuations start to decrease. The values reported in the table obviously refer to the initial time, $t=0$. Our main interest is in the evolution of magnetic fluctuations, as they decrease in time (compared to the background field), leading to a higher predicted PD. Simulations were performed using a resolution of $512^3$ grid points, employing reconstruction and derivation routines for an overall fifth order of spatial accuracy (in smooth regions without discontinuities), a third order Runge-Kutta time-stepping algorithm, and the ``upwind constrained transport'' (UCT) method of preserving a zero divergence of the magnetic field \citep{Londrillo2004,DelZanna2007}. Each simulation lasted about half an hour, using 8 Ampere A100 GPUs (each of them treating $256^3$ cells) on the Leonardo cluster.

\subsection{Properties of turbulence}


\begin{figure}[t]
    \centering
    \includegraphics[height=75mm]{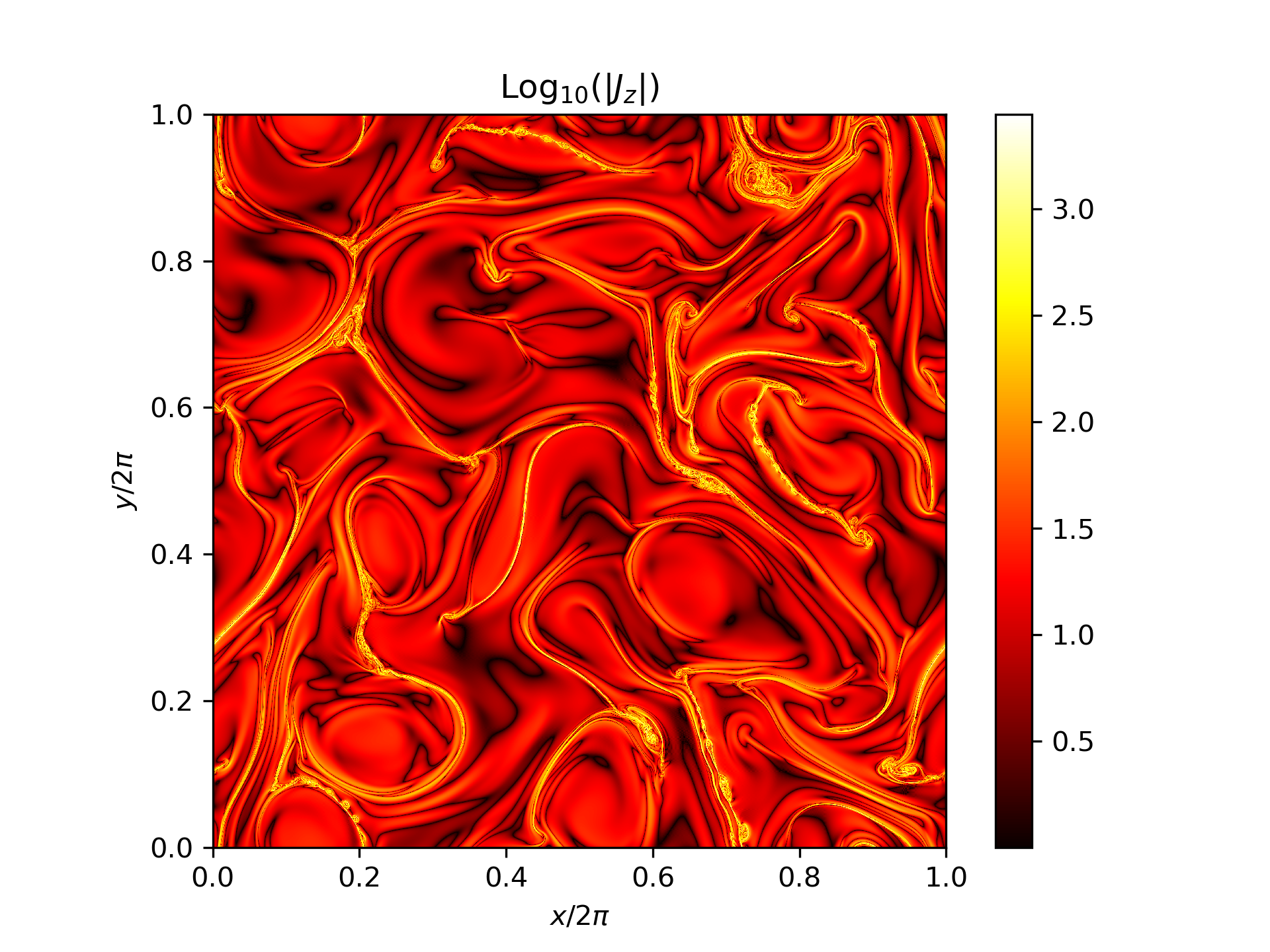}
    \caption{Typical output of relativistic MHD turbulence simulations obtained with the novel version of the \texttt{ECHO} code. Here we refer to the case described in \cite{DelZanna2024} (2D run at the resolution of $4096^2$), and we display the module of $\nabla\times\bm{B}$ parallel to the mean field.}
    \label{fig:Jz}
\end{figure}



\begin{figure*}[t]
    \centering
    \includegraphics[width=0.45\linewidth]{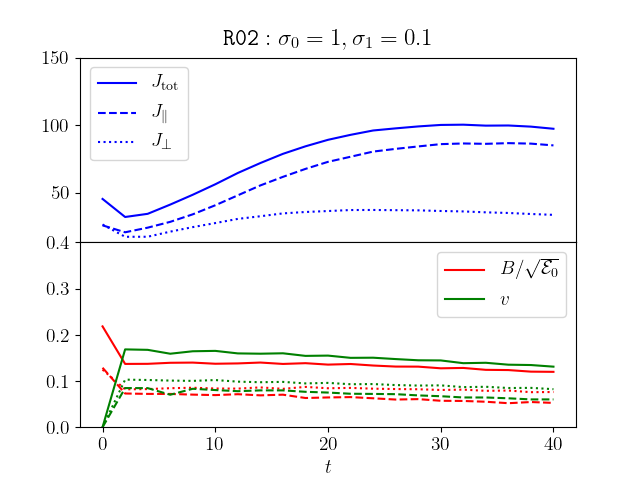}
    \includegraphics[width=0.45\linewidth]{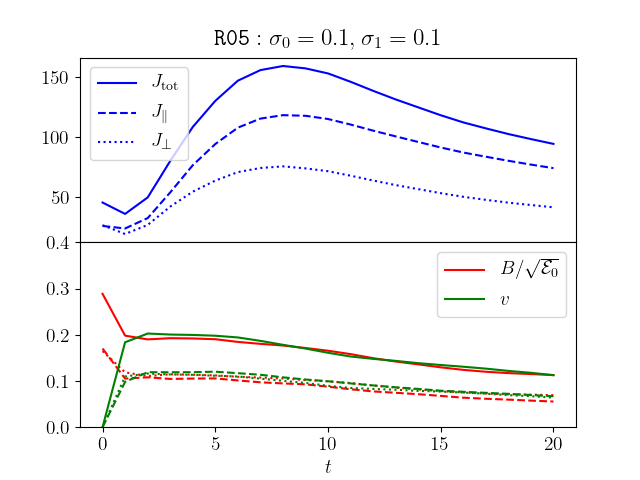}
    \includegraphics[width=0.45\linewidth]{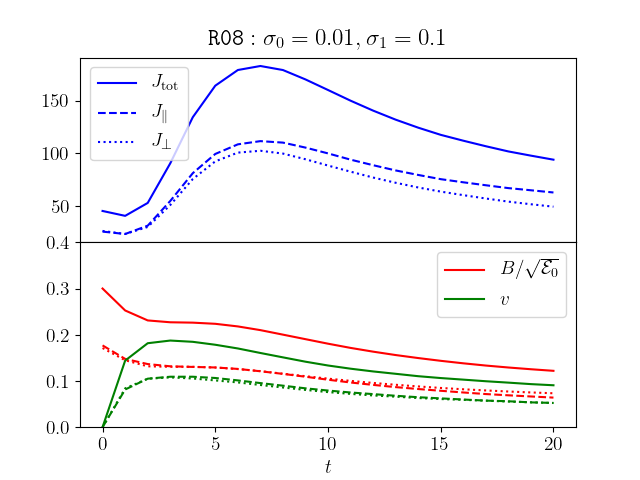}
    \includegraphics[width=0.45\linewidth]{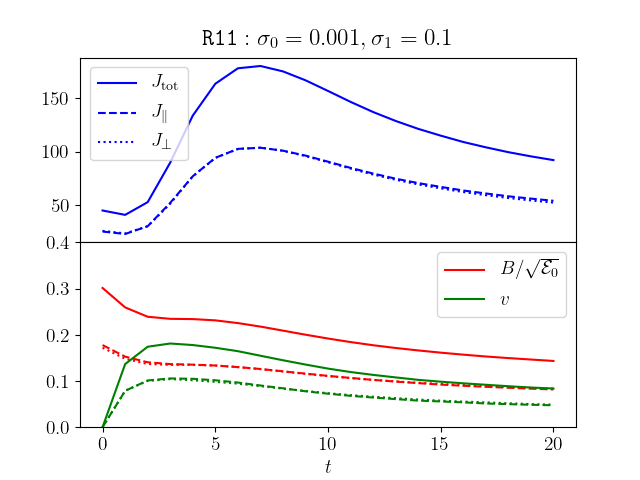}
    \caption{Time sequences of rms quantities for a selection of four runs, choosing a varying $\sigma_0$ while keeping $\sigma_1=0.1$ fixed. In the upper panels we show the total current density, $J=|\nabla\times\bm{B}|$, $J_\parallel$ (along $\bm{B}_0$), and $J_\perp$, whereas in the bottom ones we plot $B/\sqrt{\mathcal{E}_0}$, basically the Alfv\'en speed, and $v$, again total, parallel, and perpendicular components, adopting the same notations.}
    \label{fig:rms}
\end{figure*}


It is well established that MHD turbulence, including its relativistic version, is characterized by the formation of thin current sheets at small scales where most of the dissipation occurs. In \cite{DelZanna2024} it is demonstrated that the high-order algorithms of the \texttt{ECHO} code allow one to obtain extended inertial scales (two full decades in 2D high-resolution runs using $4096^2$ grid points) and a steep cascade to the dissipation scales, even in the absence of explicit dissipative terms; that is, in the ideal MHD case. In Fig.~\ref{fig:Jz} we show the module of the (nonrelativistic) current density $J_z = (\nabla\times \bm{B})_z$ at $t=10$ for the run described in that paper. This corresponds to the 2D case with $\vartheta=0$ in Eq.~(\ref{eq:setup}), using the parameters $\sigma_0 = 0.5$ (the \textit{cold} magnetization was 100) and $\eta=0.25$. Notice the fine details of current sheets, where hints of ongoing reconnection and plasmoid instabilities can be seen, as is usually found in resistive (relativistic) MHD simulations at a much higher resolution \citep{Chernoglazov2021}. 

Rather than investigating particular configurations assumed by the system, which differ from time to time or by selecting different slices of our cubic domain, we are more interested in statistical information and in the behavior of the system at different spatial scales, as it is common to do in turbulence studies. Here, rms quantities and spectra are computed on top of 3D data for all physical variables, for several output times and for all runs listed in Table~\ref{Tab:param}. In the following we highlight results of particular relevance, by selecting some of the 12 runs.

We start by analyzing the time evolution of the rms of the curl of the magnetic field, which is the classical definition of the current density when the displacement current can be neglected (Ampere's law). The maximum development of MHD turbulence roughly corresponds to the peak of that quantity, so a simulation for given parameters must be run for long enough to go beyond that threshold. Moreover, since all the simulations in the present work are initialized by prescribing magnetic fluctuations alone, the turbulent cascade will develop only after a transient phase in which velocity fluctuations are also triggered by the unbalanced magnetic configuration, as we see below. 

In the top panels of Fig.~\ref{fig:rms}, we plot time series for the total, parallel, and perpendicular components of $\bm{J} = \nabla\times \bm{B}$ (blue curves) for four simulations, with the ratio $\sigma_1/\sigma_0$ ranging from $0.1$ up to 100; that is, from a dominant guide field to a negligible one. Notice that the peak of turbulence is reached around $t\simeq 7$ for the \texttt{R08} and \texttt{R11} cases, corresponding to large normalized amplitude $\eta > 1$ values, while $t\simeq 30$ for the \texttt{R02} case (and beyond 50 for \texttt{R01}), for which $\eta < 1$.  As was expected, the anisotropy  between $J_\parallel$ (dashed lines) and $J_\perp$ (dotted lines) observed in the first selected run (case \texttt{R02}) is large, whereas it progressively reduces when the guide field is small, meaning that fluctuations dominate over the mean field and the overall behavior tends to full isotropy (case \texttt{R11}). Moreover, when the guide field is important, $J_\parallel$ is the dominant component, because it refers to magnetic fluctuations developing in the plane transverse to $\bm{B}_0$, where turbulence is expected to be stronger.

A similar trend is observed in the lower panels, where the time evolution of the \textit{rms} of velocity $\bm{v}$ (green curves) and Alfv\'en speed $\bm{B}/\sqrt{\mathcal{E}_0}$ (red curves) are plotted. Here the total inertia is computed at $t=0$, including the initial magnetic fluctuations, $\mathcal{E}_0 =\rho h + B^2 = \rho h (1 + \sigma_0 + \sigma_1)$, with $\rho h \simeq 4p = 1000$ in all our runs. Notice that velocity fluctuations are initially zero, given that only magnetic fluctuations are imposed at $t=0$, and they remain small for all runs at any time, so that relativistic corrections for synchrotron emission (such as Doppler boosting and polarization angle swing, described in the next section) are expected to be small. After the initial transient, when the velocity fluctuations increase, a sort of Alfv\'enic balance is soon reached, in which both quantities slowly decrease in time following extremely similar patterns. When the guide field is large, the Alfv\'enic coupling is so strong that the two quantities are practically equivalent (see cases \texttt{R02} and \texttt{R05}). When $\bm{B}_0$ decreases compared to fluctuations, as in cases \texttt{R08} and \texttt{R11}, the Alfv\'enic coupling is less strong and the kinetic fluctuations always remain smaller than the (normalized) magnetic one, even if trends are still similar.


\begin{figure}
    \centering
    \includegraphics[height=70mm]{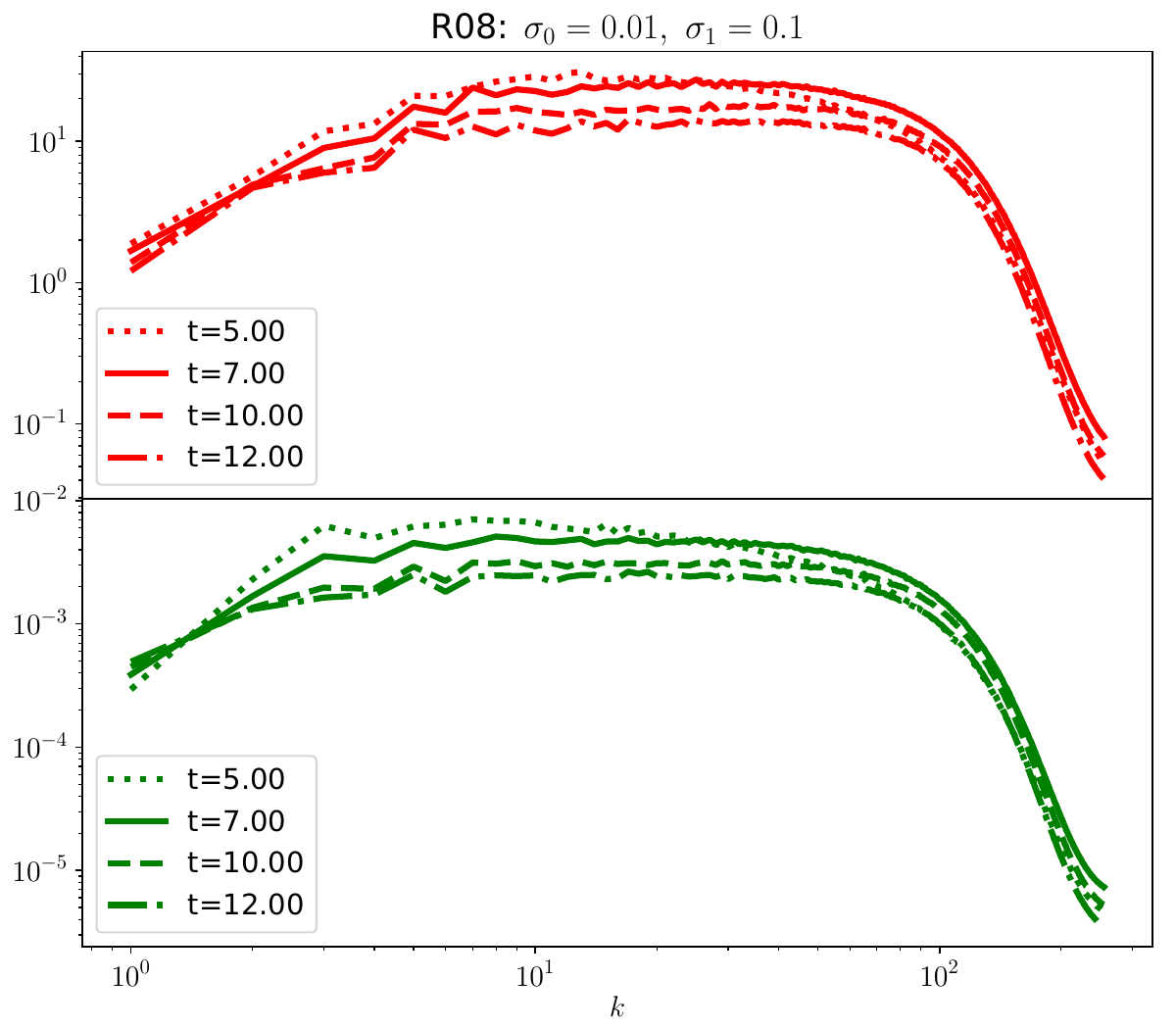}
    \caption{Omnidirectional, compensated spectra of magnetic (red) and kinetic (green) energies, for the $\texttt{R08}$ run and several output times.}
    \label{fig:spectra-time}
\end{figure}



\begin{figure*}[t]
    \centering
    \includegraphics[height=68mm]{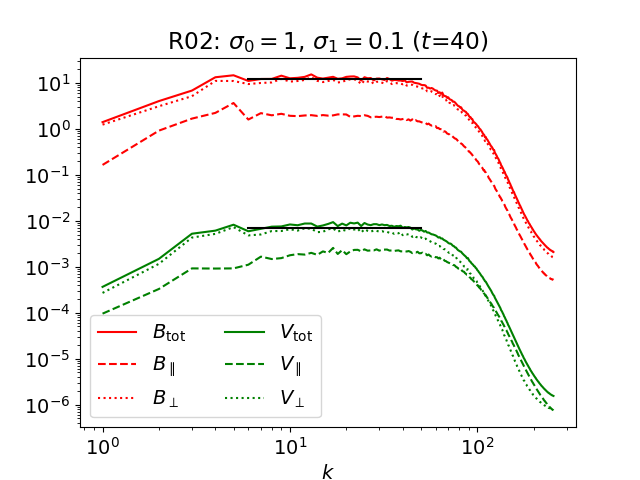}\hspace{5mm}
    \includegraphics[height=65mm]{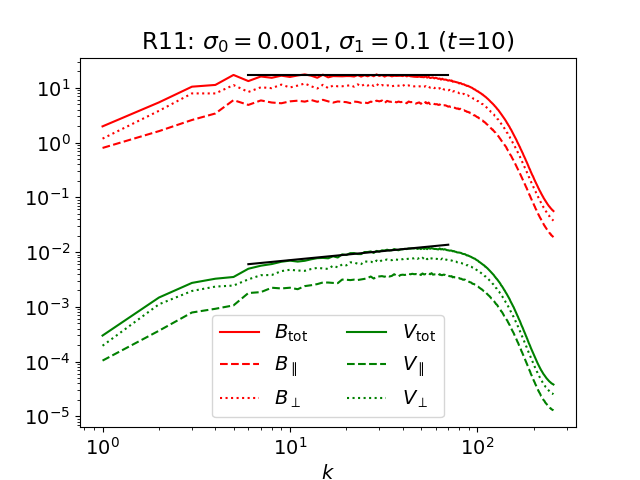}
    \caption{Omnidirectional spectra, compensated by $k^{5/3}$, of magnetic (red) and kinetic (green) energies, with total (solid), parallel (dashed), and perpendicular (dotted) components. In the left panel, we show spectra for the $\texttt{R02}$ run at $t=40$, while the case $\texttt{R11}$ at $t=10$ is shown in the right panel (both times are after the peak of turbulence, see the corresponding blue curves in Fig. \ref{fig:rms}. Horizontal black lines indicate a $k^{-5/3}$ slope, while the kinetic spectra for $\texttt{R11}$ are better fit by a $k^{-4/3}$ slope.}
    \label{fig:spectra}
\end{figure*}



\begin{figure*}
    \centering
    \includegraphics[height=65mm]{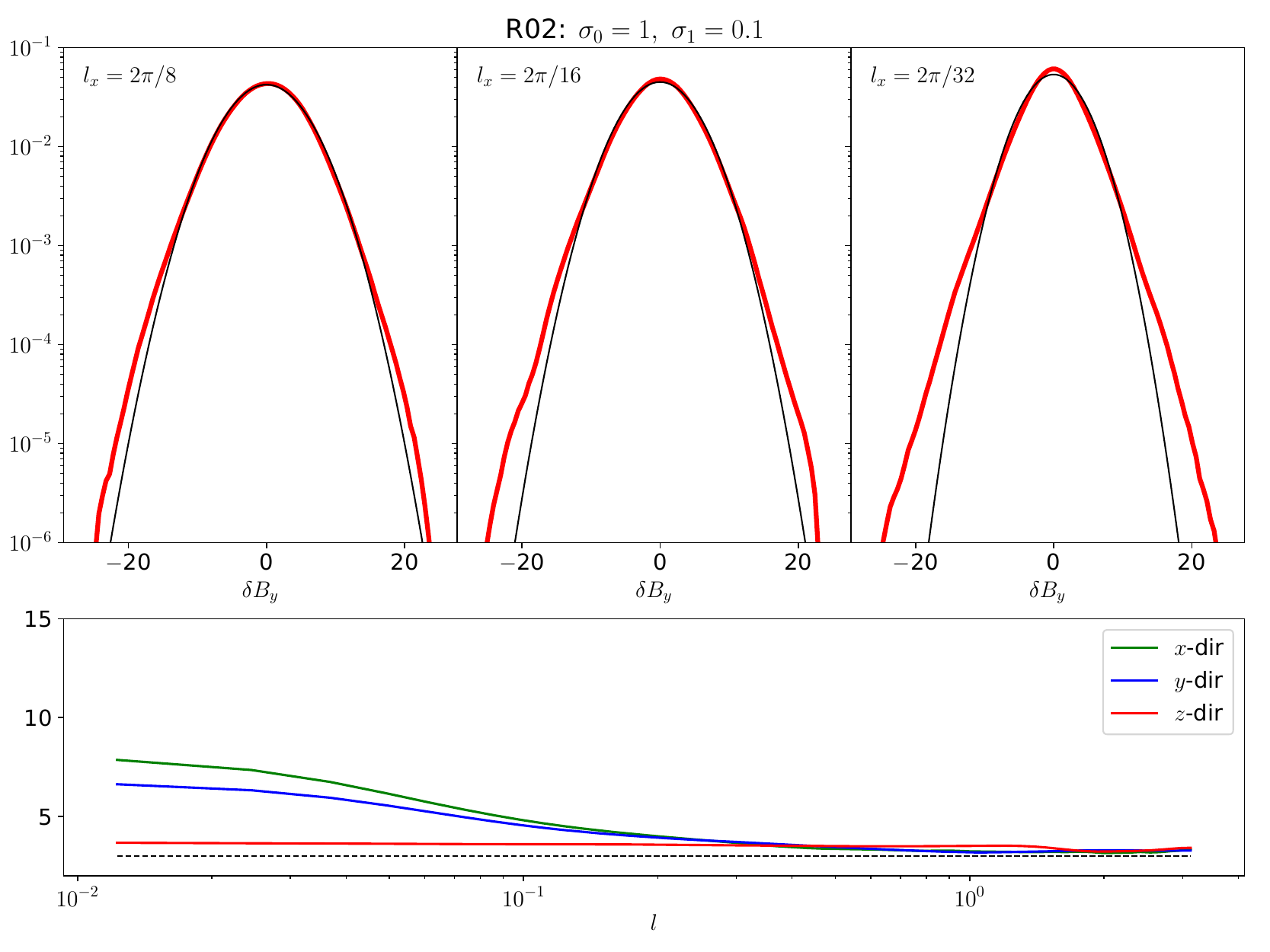}\hspace{5mm}
    \includegraphics[height=65mm]{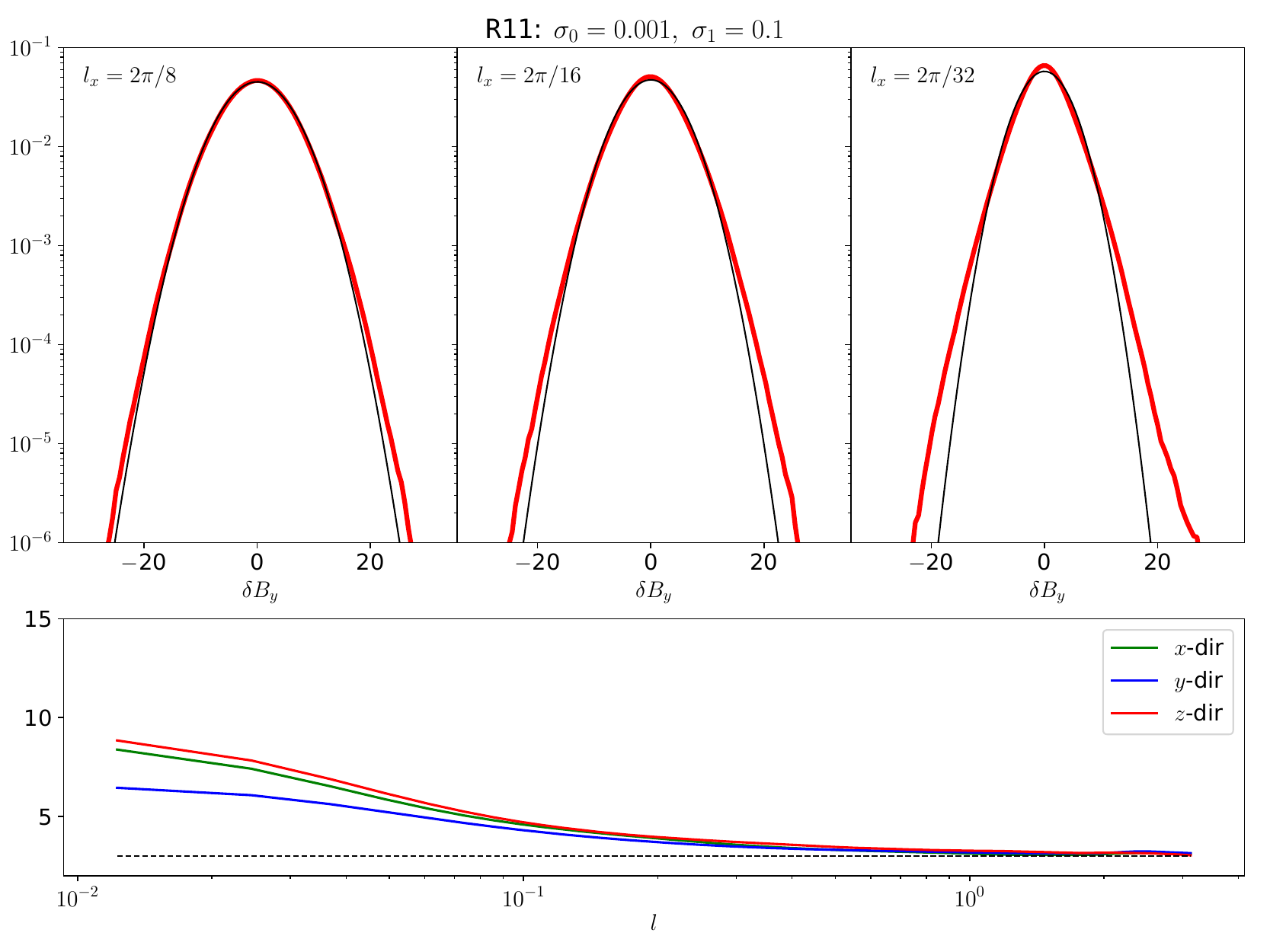}\hspace{5mm}
    \caption{Statistical properties of turbulence for the same runs and output times of Fig.~\ref{fig:spectra}. In the top panels, we plot the PDF of $\delta_x B_y$ for three different spatial separations, $l=2\pi/k$. In the bottom panels, the kurtosis, $\mathcal{K}$, is plotted against the separation scale, $l$, for$\delta_i B_y$, with $i=1,2,3$.}
    \label{fig:kurtosis}
\end{figure*}


When dealing with MHD turbulence, the spectral behavior of fluctuations must necessarily be investigated. We selected three runs with very different values of $\eta$, and show the results in Fig.~\ref{fig:spectra-time} and Fig.~\ref{fig:spectra}. The 3D power spectral densities depend on $k_x,k_y,k_z$, but these can be isotropized by summing over contributions with the same module, $k$, providing omnidirectional 1D spectra \citep[e.g.][]{Franci2018}: $E_\mathrm{m}(k)$ of $(\delta \bm{B})^2$ (red curves, not normalized against $\mathcal{E}_0$) and $E_\mathrm{v}(k)$ of $(\delta \bm{v})^2$ (green curves). These are further multiplied by $k^{5/3}$, or other slopes, to enhance similarities or deviations from a Kolmogorov spectrum $E(k)\propto k^{-5/3}$. Notice the inertial range covering more than a full decade, followed by a quite rapid decay toward the (numerical) dissipation scales. 

In Fig.~\ref{fig:spectra-time} we use spectra of run $\texttt{R08}$ to show their time dependence in a typical decaying turbulence simulation. While before the peak of turbulence the inertial range has not yet reached a stationary equilibrium, around the peak ($t\simeq 7$) and for later times the inertial ranges are fully developed and the slopes are maintained, whereas only the amplitudes of the fluctuations decay. We find that while $E_\mathrm{m}(k)$ follows a pure Kolmogorov spectrum, $E_\mathrm{v}(k)$ unexpectedly shows a less steep slope. In order to investigate this behavior better, in Fig.~\ref{fig:spectra} we report the power spectral energies from two opposite cases in terms of the parameter $\eta$: $\texttt{R02}$ (strong mean field compared to fluctuations) and $\texttt{R11}$ (fluctuations dominate, hence we expect a more isotropic situation). Other than just the total spectra (solid lines), here we also show the separate contribution from parallel (dashed) and perpendicular fluctuations (dotted): the latter always dominate, being twice the parallel contribution in the almost isotropic case of $\texttt{R11}$, whereas in the anisotropic case when $\bm{B}_0$ of $\texttt{R02}$, the parallel component is negligible. As far as the spectral slope is concerned, the (total) magnetic energy, $E_\mathrm{m}(k)$,  invariably shows a pure Kolmogorov spectrum (all spectra in the figure are compensated for by $k^{5/3}$), for all our 12 runs. On the other hand, the kinetic energy spectra, $E_\mathrm{v}(k)$, are Kolmogorov-like when $\eta$ is small, as in $\texttt{R02}$, or rather decay less steeply, roughly as $k^{-4/3}$ (see the green curves in Fig.~\ref{fig:spectra}), as was already shown in Fig.~\ref{fig:spectra-time}. We believe that this could be due somehow to the lack of an exact Alfvénic balance, as is observed in Fig.~\ref{fig:rms} for runs with a high $\eta$, so that velocity fluctuations are not coupled to the behavior of magnetic ones.

Other than the precise value of the slope in the inertial range of the spectra of various quantities, it is interesting to investigate its statistical properties and departures from Gaussianity, typical of any realistic turbulent cascade, where correlations and formation of structures are expected. Let $\mathcal{P}(\delta_i B_j; l)$ be the (normalized) probability distribution function (PDF) for the quantity
\be
\delta_i B_j = B_j(x_i) - B_j(x_i - l),
\ee
where $i,j$ can assume the values $1,2,3$ indicating the spatial directions, and $l$ is a parameter indicating a given spatial separation scale. Here we define $z$ as the coordinate along the parallel direction, so that $x-y$ will be the perpendicular plane, where the turbulent properties may be different, especially for runs with a dominant guide field, $\bm{B}_0$, (large $\sigma_0/\sigma_1$ ratios). We then define the moment of order $n$ for such a quantity as
\be
\mathcal{S}_n = \int_{-\infty}^{+\infty} |\delta_i B_j|^n \, \mathcal{P}(\delta_i B_j; l_i) \,\mathrm{d} \delta_i B_j, 
\ee
and the ``kurtosis'' parameter, defined to be the normalized moment of order $n=4$; that is
\be
\mathcal{K} = \frac{\mathcal{S}_4}{(\mathcal{S}_2)^2},
\ee
where the denominator is basically the square of the variance, or equivalently the fourth power of the standard deviation, and all the above statistical functions are clearly function of the scale separation, $l$. When fluctuations are completely uncorrelated and randomly distributed, and hence $\mathcal{P}(\delta_i B_j; l)$ coincides with the Gaussian (normal) PDF, then $\mathcal{K}=3$ at any scale, $l$, whereas in a well-behaved turbulence one would expect higher values of the kurtosis at small scales (``intermittency'').

Selecting the same runs and output times of Fig.~\ref{fig:spectra}, in Fig.~\ref{fig:kurtosis} we report, in the top panels, the PDFs for $\delta_x B_y$ at three spatial scales, $l=2\pi/k$, within the inertial range, with the wave number ranging from $k=8$ to $k=32$. The choice of directions should highlight the formation of structures (vortexes or current sheets) in the perpendicular plane, with variations in one component of $\bm{B}$ in the transverse direction. We have verified that, remaining in the perpendicular plane $x-y$, the behavior for $\delta_y B_x$ is very similar, as was expected. As we can see, the smaller the spatial separation scale, the stronger the departure from a Gaussian distribution, meaning that more events are statistically expected in the tails of the distributions. In the lower panels, we plot the kurtosis, $\mathcal{K}$, as a function of the separation scale, $l$, for $\delta_x B_y$ (green line), $\delta_y B_y$ (blue line), and $\delta_z B_y$ (red line). Notice how strong departures from Gaussianity (corresponding to the dashed line with $\mathcal{K}=3$) are observed at small scales, especially below values around $l=2\pi/32 \simeq 0.2$ of the above panels, toward the end of the inertial range in $k$ space.

Important differences arise depending on the run parameters. The plots displayed on the left panels are for \texttt{R02}, where the guide mean field is the strongest ($\sigma_0=1$). In this case we observe uncorrelated fluctuations of $B_y$ (and for the other components is the same) in the parallel direction, $z$, so that the kurtosis remains $\mathcal{K}\approx3$ at all scales (we recall that the corresponding spectrum was also weaker compared to that of the perpendicular components; see the dashed red lines in Fig.~\ref{fig:spectra}, left panel). On the contrary, when the mean field is negligible as in \texttt{R11} ($\sigma_0=0.001$), in the right panels, the kurtosis in the $z$ direction behaves basically like that for $x$, since both directions are normal to the direction of $B_y$ and basically indistinguishable (the red curves tends to coincide with the green one, and it is well above the horizontal dashed black line with $\mathcal{K}=3$).


\begin{figure}[t]
    \centering
    \includegraphics[height=65mm]{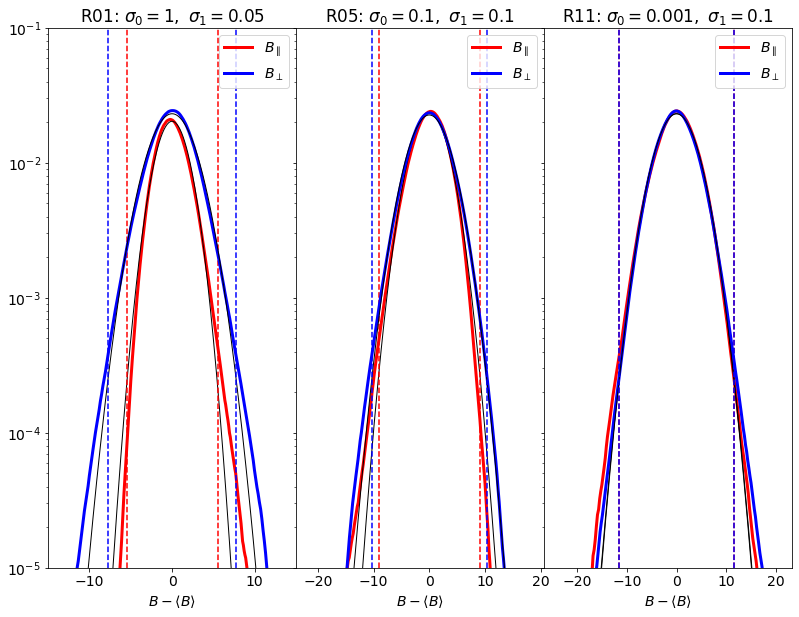}
    \caption{Probability distribution functions (PDFs) for magnetic field parallel (red lines) and perpendicular (blue lines) components, for three runs with increasing $\sigma_1/\sigma_0$ (and $\eta = B_1/B_0$). The overplotted black lines refer to Gaussian normal distributions with the same standard deviation as the corresponding PDF, and the dashed vertical lines indicate the positions at three standard deviations ($99.73$ probability).}
    \label{fig:anisotropy}
\end{figure}


We conclude this section by analyzing the PDFs of magnetic components themselves, given that, as discussed in the introduction, we want to compare our results against analytical estimates for the PD, which are strictly valid in the case of Gaussian stochastic distributions, with the same variance for all components of the vector $\delta {\bm B}$ (isotropic case). In Fig.~\ref{fig:anisotropy} we plot the PDFs for both the parallel ($B_\parallel = B_z$, red lines) and perpendicular ($B_\perp$, average of the PDFs of $B_x$ and $B_y$, blu lines) for three simulations with the increasing ratio $\sigma_1/\sigma_0$. When this is small (0.05 for \texttt{R01}, corresponding to $\eta=B_1/B_0 = 0.224$, on the left hand side) the anisotropy is rather strong, with a much broader distribution for $B_\perp$, and deviations from Gaussianity in the tails of both distributions (intermittency). When the ratio is large (100 for \texttt{R11}, $\eta = 10$, on the right hand side), the situation is the opposite: the guide field is basically negligible with respect to fluctuations, so the latter are isotropic and substantially Gaussian, with very little intermittency. The case \texttt{R05}, with $\sigma_1 = \sigma_0$ ($\eta=1$), is clearly intermediate between the other two. Variance anisotropy is a well-known property of any turbulent plasma in the presence of a mean guide field. It is either observed in solar wind data \citep{Bavassano1982} or in 3D MHD simulations \citep{Matthaeus1996}.

\section{Synchrotron emission and polarization degree}

In the present work, we model any synchrotron emission source in the simplest possible way: a cubic box of plasma with a (constant) main magnetic field, $\bm{B}_0$, embedded in an evolving turbulent plasma, filled by a power law of emitting electrons, leading to an emissivity of the form $j_\nu\propto\nu^{-\alpha}$ (here we assume $\alpha=0.6$). 
The recipes for computing synthetic synchrotron emission and polarization properties on top of (ideal) relativistic MHD simulations were given and explained in \cite{DelZanna2006}, in which Doppler boosting and relativistic transformation rules from the frame comoving locally with the fluid to that of the observer were taken into account. After integrating along the LOS one is able to obtain maps in the plane of the sky of the intensity $I$, of the Stokes parameters $Q$ and $U$ ($V=0$ for linear polarization). Then, the quantity
\be
\Pi = \frac{\sqrt{Q^2 + U^2}}{I},
\ee
is the linear degree of polarization. If integrated over the plane of the sky, it provides the PD value. For the sake of simplicity, only two limiting cases are considered here:
\begin{itemize}
    \item guide field parallel to LOS  ($\bm{B}_0 \parallel \bm{n}$).   
    When $\bm{B}_0$ is along the LOS, we expect a weak polarized intensity and PD. The limiting case for very small fluctuations is that of a complete unpolarized emission with $\Pi=0$.
    \item guide field perpendicular to LOS  ($\bm{B}_0 \perp \bm{n}$).
    When $\bm{B}_0$ is in the plane of the sky we expect a stronger polarized intensity and a higher PD. If the magnetic field is uniform, one should obviously expect to find $\Pi = \Pi_\mathrm{max}\simeq 70\%$.

\end{itemize}

\subsection{Predicted and computed global PD}

When a parcel of plasma can be considered as static (hence neglecting velocity effects such as Doppler boosting or polarizations angle swing), and uniform (assuming that the density of emitting particles is also constant), simple analytical predictions for the Stokes parameters and for the PD can be made even in the presence of magnetic fluctuations. In \cite{Bandiera2016} the situation characterized by a mean background field, $\bm{B}_0$, and random fluctuations, $\delta\bm{B}$, with \emph{isotropic} Gaussian PDFs (i.e., with the same standard deviation, $\sigma$, for each component), was studied and, in particular, an analytical prediction for the total PD was proposed. This was shown to be a function of $\bar{B}^2/(2\sigma^2)$ alone, where $\bar{B}$ is the projection of the background field on the plane of the sky and $\sigma^2$ the variance of magnetic components, the same for all directions. In the present section, we only consider the case of a guide field perpendicular to LOS  ($\bm{B}_0 \perp \bm{n}$), so that $\bar{B}\equiv B_0$ (otherwise, the sine of the inclination angle should appear), and we prefer to use the total intensity of fluctuations, $\delta B^2$ (say the rms value, squared), with $\delta B^2=3\sigma^2$ in the isotropic case. Hence, the ratio $(\delta B/B_0)^2$ directly coincides with the ratio of magnetic energies, $\delta E/E$, in the disordered and ordered components \citep{Bucciantini2017}.

Using our definitions, the analytical function by \cite{Bandiera2016} for the global PD can be rewritten as (BP16 model from now on)
\be
\label{eq:BP16}
\frac{\Pi}{\Pi_\mathrm{max}} = \frac{3+\alpha}{4} \,\xi \,\frac{{}_1 F_1[\,(1-\alpha)/2, 3; - \xi \,]}{{}_1 F_1[\, - (1+\alpha)/2, 1; - \xi \,]}, \quad
\xi = \frac{3}{2}\left( \frac{\bar{B}}{\delta B} \right)^{2}\!\!,
\ee
where ${}_1 F_1[a, b; z]$ is the Kummer's function, or confluent hypergeometric function of the first kind, and $\alpha$ is the usual spectral index. The above formula was actually derived 50 years before by \cite{Burn1966}, in a slightly different but equivalent form, as one can easily check by using the Kummer's transformation ${}_1 F_1[a, b; z] = {}_1 F_1[b-a, b; -z]\, \mathrm{e}^z$.

\begin{figure}
    \centering
    \includegraphics[height=65mm]{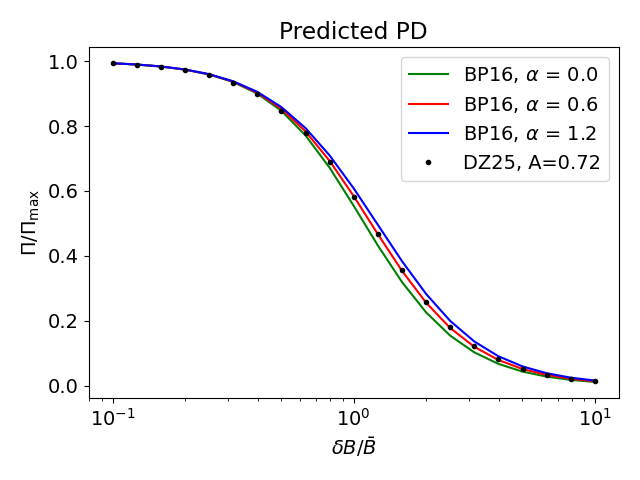}
    \caption{Comparison of the predicted PD for isotropic Gaussian fluctuations based on confluent hypergeometric functions by \cite{Bandiera2016} against our simpler formula in Eq.~(\ref{eq:DZ25}). The expression for $A=0.72$ (black dots) fits the BP16 model corresponding to $\alpha=0.6$ (the value adopted here, the central red curve) extremely well.}
    \label{fig:analytical}
\end{figure}

In Fig. \ref{fig:analytical} the BP16 analytical estimate is plotted for three different values of the spectral index $\alpha$, ranging from 0 to 1.2, in a wide range of values of $\delta B/\bar{B}$. The central red curve for $\alpha=0.6$, the value assumed in the present work, is compared against a much simpler formula for the total PD, proposed here for the first time (DZ25 model, from now on):
\be
\label{eq:DZ25}
\frac{\Pi}{\Pi_\mathrm{max}} = \left[1+ A \left( \frac{\delta B}{\bar{B}} \right)^2\right]^{-1},
\ee
where $A$ is a free numerical parameter. The above expression depends in a simple way on $\delta B/\bar{B}$, with obvious asymptotes 1 and 0, respectively, for negligible and large fluctuations, as was expected. The value of the parameter - we chose $A=0.72$ - is simply derived from the best fit of the above functional form against the BP16 model for $\alpha=0.6$. As was already shown in the BP16 paper, we can see here too how curves are very similar for the chosen values of $\alpha$, all in a reasonable range for the sources of interest. For the particular value $\alpha = 1$ (power law index $p=3$), the predicted normalized PD in Eq.~(\ref{eq:BP16}) reduces \emph{exactly} to $\xi / (1 + \xi)$; that is, to Eq.~(\ref{eq:DZ25}) with $A=2/3$ \citep{Burn1966}. Given the mild dependence on $\alpha$, even this value may be considered to yield a good approximation; however, in the remainder of this work, we prefer to use the simple analytical formula of Eq.~(\ref{eq:DZ25}) with $A=0.72$.

As we have seen in the previous section, our turbulence runs do not fit the simplifying conditions for the above analytical predictions. It is true that the velocity always remains mildly relativistic, but magnetic fluctuations are not Gaussian at all scales, and, above all, in the presence of a significant mean field, $\bm{B}_0$, they are not isotropic, with different values of the standard deviation in the parallel and perpendicular directions. It is thus interesting to quantify possible departures of the computed mean PD (for each run and at every output time) from the analytically estimated one.

\begin{figure}
    \centering
    \includegraphics[height=65mm]{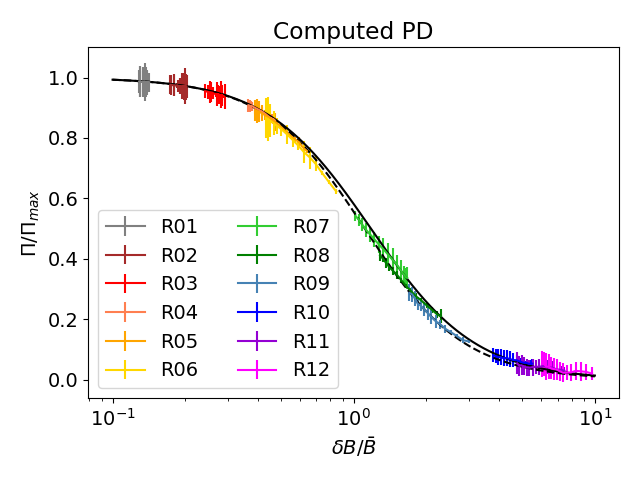}
    \caption{Mean PD computed assuming a guide field in the plane of the sky and perpendicular to LOS  ($\bm{B}_0 \perp \bm{n}$). Results for all 12 runs are shown as function of $\delta B/\bar{B}$, from left to right for increasing values of $\eta$, while time increases from right to left, one tick per output. The results basically all lie, within the error bars (estimated from the polarization along $\bm{B}_0$), on the solid black line, corresponding to the DZ25 model with $A=0.72$, as shown in Fig.~\ref{fig:analytical}. The dashed black line is an analytical improved estimate, basically obtained by increasing $A$ of $\approx 15\%$, to take anisotropy of fluctuations into account (see the text and Appendix \ref{app:nicco-true-model}).}
    \label{fig:Rino-plot}
\end{figure}

In Fig.~\ref{fig:Rino-plot} we show precisely this comparison. Each color indicates a different run, for increasing values of $\eta$ (the initial $\delta B/\bar{B}$ value) from left to right. Output times (the ticks, 15 for each run) increase instead from right to left, since we simulate a decaying turbulence. Only output results for times after the peak of turbulence are considered, with a separation of $\Delta t =2$ for the cases with $\sigma_0=1$, showing a slower evolution, and $\Delta t =1$ otherwise. We can clearly see that only very small deviations from the theoretical estimate (the solid black line) are apparent, and these are concentrated at the center of the plot, for runs where $\delta B \simeq \bar{B}$ or more. The error bars correspond to the averaged polarized intensity along the mean magnetic field, that is along the LOS, taken as a measure of intrinsic stochasticity. For the analytical estimate, we used the DZ25 fit with $A=0.72$, which agrees very well with the BP16 exact model, as is demonstrated in Fig.~\ref{fig:analytical}.  

In the same cited paper \citep{Bandiera2016}, whose results have recently been extended to particle distribution functions different from a simple power law \citep{Bandiera2024}, it is also shown that it is possible to account for the effects of anisotropic magnetic fluctuations. Unfortunately, in the presence of a nonzero mean background  magnetic field, there is no closed analytical formula to estimate the global PD, unlike the isotropic case. However, as long as the level of anisotropy is not too strong, one can derive an approximated relation, introducing corrections to Eq.~(\ref{eq:BP16}) (see Appendix~\ref{app:nicco-true-model} for further details). Analogously, it is also possible to show, through a simple heuristic argument, how the approximated relation Eq.~(\ref{eq:DZ25}) can be modified, to account for this anisotropy. One just needs to recall that only the perpendicular fluctuations, $\delta B_\perp$ (in the plane of the sky), contribute to depolarization. In the presence of anisotropy, as was discussed in the previous section and as is apparent from Fig.~\ref{fig:anisotropy}, we have $\delta B_\perp > \delta B_\parallel$; hence, one can deduce $\delta B_\perp^2 > \delta B^2/3$, where we recall that the latter term represents the averaged variance over all directions (coincident to that for every direction in the isotropic case). At the same time, an effective mean magnetic field is expected to decrease with respect to $B_0$ ($\equiv \bar{B}$ in our case), given that fluctuations in the parallel direction are smaller in the anisotropic case (Fig.~\ref{fig:anisotropy}). This combined effect can thus be simply modeled as an increase in the value of the parameter $A$ in Eq.~(\ref{eq:DZ25}). Indeed, as is shown in Fig.~\ref{fig:Rino-plot}, raising $A$ from 0.72 to an effective value of 0.85 (dashed line) provides a tight bound to our results, which now fits all simulation outputs much better, especially in the central region of the plot where modifications were mostly needed. This  suggests that the discrepancy with the previous prediction (the solid line curve) is really due to the presence of anisotropic PDFs in our simulated turbulent plasma. For further details, including a measure of the anisotropy degree in our runs, see  Appendix~\ref{app:nicco-true-model}.

\subsection{Statistics of synchrotron emission}

Observations of synchrotron emission and its polarization properties are powerful probes of the magnetic geometry in a source. This is certainly true for uniform or smoothly varying magnetic fields, but when turbulence is present the statistics of synchrotron polarization maps is also affected, so that information on turbulence properties can in principle be drawn by observational data, as was already discussed in relation to the global PD in the previous subsection. Given what was discussed in section 2, we also expect intermittency and deviations from isotropy to play a role. Here we analyze the synchrotron maps and statistical properties computed on top of three selected runs with very different values of the normalized initial amplitude $\eta = B_1/B_0 = \sqrt{\sigma_1/\sigma_0}$, starting from $\eta = 0.224$ of \texttt{R01}, then $\eta = 1.41$ of \texttt{R06}, to $\eta = 14.1$ of \texttt{R12}, computing the synthetic emission properties at output times after the respective peak of turbulence. This is done for the two scenarios of a mean magnetic field parallel to the LOS (e.g., when a PWN torus threaded by a toroidal field is observed at the limb), and when the mean magnetic field lies in the plane of the sky (the center of the torus), leading to very different results.

\begin{figure*}[h]
    \centering
     \includegraphics[height=18cm]{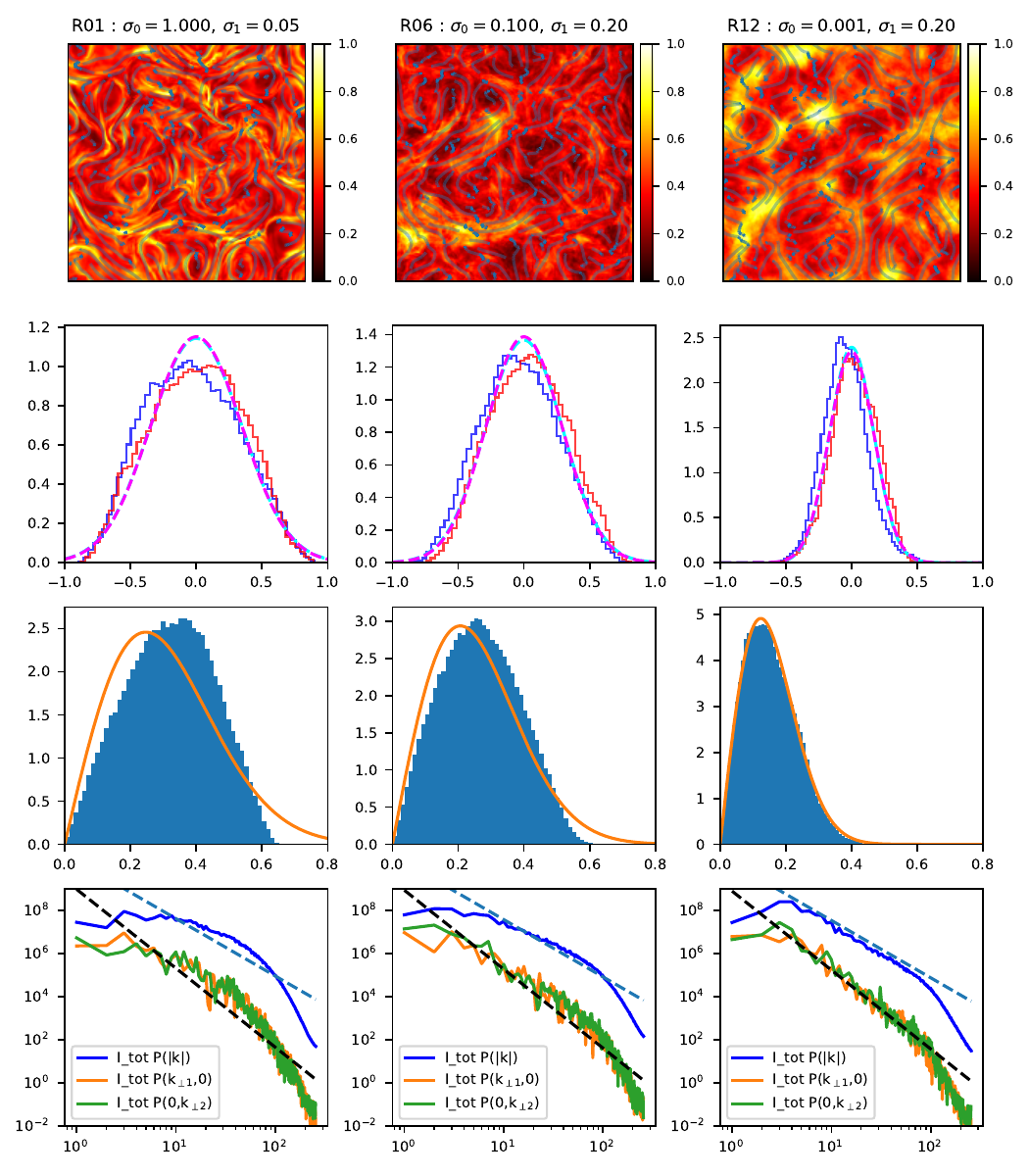}
    \caption{Synchrotron properties for the case of a mean field along the LOS ($\bm{B}_0 \parallel \bm{n}$), for cases \texttt{R01} ($t=60$), \texttt{R06} ($t=10$), and \texttt{R12}, ($t=10$). Upper row: Total synchrotron intensity normalized to the maximum, overlaid with the synchrotron magnetic fieldlines (cyan curves) estimated from Stokes parameters. Second row: Distribution function of Stokes parameters $Q/I$ (blue) and $U/I$ (red) together with $0$-mean Gaussian distribution with the same variance (cyan and magenta respectively), normalized to $\Pi_{\rm max}$. Third Row: Distribution function of the PD, together with the Rice function with variance derived from Stokes. Fourth row: Spectral properties of the total intensity, compared with the theoretical expectation for Kolmogorov turbulence both on-shell $\propto k^{-8/3}$ (dashed blue) and in the $x$ and $y$ directions in $k$ space $\propto k^{-11/3}$ (dashed black).}
    \label{fig:polz}
\end{figure*}

\begin{figure*}
    \centering
     \includegraphics[height=18cm]{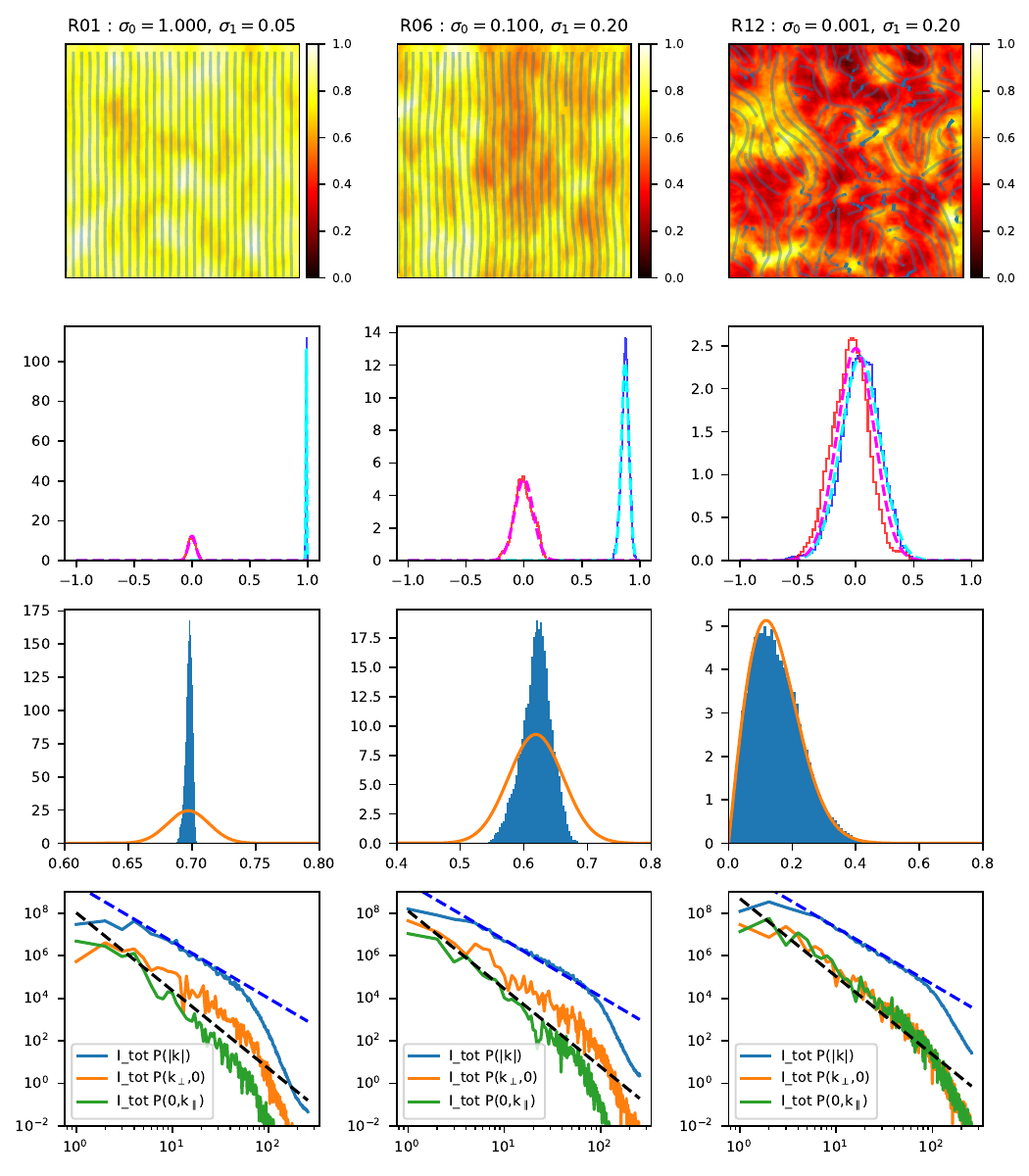}
    \caption{Same as in Fig.~\ref{fig:polz}, here for the case of a mean field contained in the plane of the sky ($\bm{B}_0 \perp \bm{n}$).}
    \label{fig:polx}
\end{figure*}

In Fig.~\ref{fig:polz} we show the polarization and emission properties for $\bm{B}_0 \parallel \bm{n}$ - that is when there is no mean field projected on the plane of the sky - for the three selected runs with varying values of $\eta$. Even if in all cases both the total and polarized intensities would vanish for a uniform field, it is immediately evident that the different levels of fluctuations and intermittency have clear effects. For small values of $\eta$, the total emission integrated along the LOS shows the presence of thin elongated and coherent emission structures that likely trace similar underlying magnetic configurations, a clear consequence of a high degree of intermittency, always present when the guide field dominates over fluctuations (see section 2). On the other hand, at high $\eta$ values, the local emissivity is characterized by a far more Gaussian incoherent pattern, and this agrees with our previous results on the statistics of turbulent fluctuations, more isotropic and Gaussian for such runs. 

The same type of result is evident by looking at the distribution of the normalized Stokes parameters, $Q/I$ and $U/I$, which have a more Gaussian-looking distribution in runs with high $\eta$ values, such as \texttt{R12}, and, more importantly, their variance is only half that of the corresponding ones in the \texttt{R01} or \texttt{R06} runs. As a result, in the first two cases the distribution of the PD values does not follow a Rice PDF as expected \citep[e.g.][]{Simmons1985}, but tends to be strongly skewed toward higher values of PD with a positive bias that is about twice the Ricean one. On the other hand, the theoretical expectation is very well reproduced for the \texttt{R12} run, and in general for runs with a large $\eta$. We recall that the width of the Stokes distribution is related to the injection spectrum, and that injections that extend to higher $|\bm{k}|$ values produce more peaked distributions. In this regard, the amount of positive bias is a measure of the coherence scale of the turbulent magnetic field with respect to the size of the emitting region (here about $L/4=0.25$). 

In the bottom panels of Fig.~\ref{fig:polz}, we show the Fourier spectrum of the map of the total synchrotron intensity, either the omnidirectional (2D) spectrum function of $|\bm{k}|\equiv |\bm{k}_\perp|$ and the 1D spectra of the two cuts in the plane of the sky. It is evident that only for isotropic Gaussian statistics of the magnetic fluctuations - that is for the \texttt{R12} run in the third column - the 2D spectrum follows $k^{-8/3}$, and correspondingly spectra $\propto k^{-11/3}$ are observed for 1D cuts in the two perpendicular directions in $k$ space, as is expected for a Kolmogorov dependence. This has been discussed in \cite{Lazarian2012} (see their Fig. 4); basically, the nonlinear dependence of emissivity with the magnetic field (squared) is weak (it would be linear only for $\alpha=1$, or $p=3$), so that spectral properties of magnetic turbulence readily reflect on statistics of emission maps, and a Kolmogorov 3D omnidirectional spectrum provides precisely $k^{-11/3}$ 1D cuts in 3D $k$ space, and $k^{-8/3}$ in a 2D $k$ space. On the other hand, when anisotropy is present, for runs with $\eta\lesssim 1$, spectra become harder, especially at scales where intermittency sets in,  providing little evidence of the presence of a well defined inertial range. This suggests that the spectrum of the synchrotron intensity, is not, \emph{per se}, a robust indicator of the underlying energy spectrum of magnetic turbulence, and that coherence and intermittency can manifest themselves in clear deviations from the standard Gaussian assumption.

In Fig.~\ref{fig:polx} the same analysis is done in the case $\bm{B}_0\perp \bm{n}$, when the mean magnetic field lies in the plane of the sky (vertically in our intensity plots). Obviously, the degree of polarization and the amount of fluctuations, both in the map of PD and in those of the Stokes parameters, are strongly dependent on the ratio of the magnetic turbulent energy over the magnetic energy associated with the mean field. An interesting finding is that, in the more anisotropic cases of \texttt{R01} and \texttt{R06}, even if the fluctuations of the normalized Stokes parameters seem to be roughly approximated by a Gaussian distribution, the variance of $Q/I$ turns out to be smaller than that of $U/I$, while in the previous analysis the two spectra were basically coincident. As a result, the distribution of the total PD strongly deviates from a Ricean profile, becoming more skewed, with values tending to pile up toward the theoretical maximum of $\Pi_\mathrm{max}\simeq 70\%$, and the typical variance being about half of what is expected  (the orange line). This trend clearly shows that the standard Ricean assumption on polarization measures might not be fully justified, while it may depend on the properties of the underlying magnetic turbulence. On the other hand, the spectra of the synchrotron fluctuation more closely follow the theoretical expectation of a Kolmogorov power law, even if the level of anisotropy by about one order of magnitude is evident between parallel and perpendicular spectra (green and orange lines, respectively). To conclude, when the mean field is negligible, which it is in the \texttt{R12} case (third column in the two figures), the results are basically very similar for both situations of $\bm{B}_0$, as was expected. Hence, in Fig.~\ref{fig:polx} too, we clearly find incoherent patterns in the map of the total intensity, nearly identical Gaussian shapes for the normalized Stokes parameters, $Q/I$ and $U/I$, a very well reproduced Rice function for the PD probability distribution, peaking at $\Pi \simeq 15\%$, the expected reduced 2D spectrum $\propto k^{-8/3}$ for the total intensity, and the now coincident 1D cuts in the two directions in $k$ space going as $\propto k^{-11/3}$.

\section{Conclusions}

In the present paper we have studied the statistical properties of turbulence in relativistically hot plasmas, and of the associated synchrotron emission, by means of numerical 3D relativistic MHD simulations, to our knowledge for the first time. Typical astrophysical sources that can be modeled using this approach are SNRs and, especially, PWNe such as the well-studied Crab nebula. Polarization observations, especially in the radio band, have always been a powerful probe to investigate the magnetic field geometry and turbulent properties in such sources. However, recently the X-ray polarimeter IXPE has obtained important new results, showing that the distribution of the linear PD is not at all uniform in these sources, possibly indicating varying levels of depolarizing turbulence in different zones \citep{Bucciantini2023,Wong2023,Liu2023,Deng2024}.

To reproduce these results, we have run 12 numerical simulations at various levels of the (relativistic) magnetization, $\sigma$, in terms of both the mean background field, $\bm{B}_0$, and stochastic fluctuations, $\delta \bm{B}$, mainly parametrized here by the ratio $\eta = B_1/B_0 = \sqrt{\sigma_1/\sigma_0}$, where $B_1 = |\delta \bm{B}|$ at the initial time. Magnetic fluctuations are initialized at large scales with random directions with respect to a background field, then the turbulent cascade develops, peaks and decays in time, and the statistical properties of the plasma and of the synthetic synchrotron emission maps are analyzed. A Kolmogorov cascade with slope $-5/3$ is invariably found for magnetic fluctuations (the velocity may instead behave differently in runs with a nearly vanishing $\sigma_0$), and turbulence rapidly relaxes to a sort of Alfvénic balance between the two components. Dissipation occurs in thin current sheets, at scales where intermittency sets in as verified by values of the kurtosis well above the canonical $\mathcal{K}=3$.

Synthetic synchrotron maps of total intensity, Stokes parameters, and PD are very different for the various runs and for the two analyzed cases of $\bm{B}_0$ along the LOS ($\Pi =0$ in the absence of fluctuations) or in the plane of the sky (allowing for the maximum PD if $\bm{B}_0$ dominates). When $\eta\gg 1$ fluctuations are predominant and the two cases show very similar properties, with incoherent patterns in the map of the total intensity, nearly Gaussian shapes for Stokes parameters, a well-reproduced Rice function for the PD probability distribution, peaking at $\Pi \simeq 15\%$, and the expected 2D reduced spectrum of $\propto k^{-8/3}$ for the polarized intensity \citep{Lazarian2012}. As far as the mean PD is concerned, in the case of $\bm{B}_0$ contained in the plane of the sky (for instance as expected at the center of the torus in PWNe), a novel and very simple analytical estimate depending on $\delta B/B_0$ alone is provided, together with a (slight) correction to take into account the anisotropic behavior of magnetic fluctuations ($\delta B_\perp > \delta B_\parallel$). The proposed function, containing a single free parameter (depending on the spectral index and on the degree of anisotropy),  agrees perfectly with the results of all our 12 runs of decaying relativistic turbulence, basically covering all possible values of the PD observed. The success of the fit of the analytical curve to data obtained from our simulations confirms that non-Gaussianity and intermittency in the PDFs occur only at small scales, basically not affecting the global synchrotron emission properties.

Other important synchrotron-emitting sources where relativistic turbulence is invariably present are certainly jets from AGNs and GRBs \citep[e.g.][and references therein]{Mattia2023,Mattia2024}, and X-ray polarimetry is again a powerful probe for the plasma conditions \citep{Tavecchio2021,Bolis2024}.  Given that the plasma velocities found in our simulations are just mildly relativistic (even letting $\bm{v}=0$ when computing synthetic emission maps, these look basically identical), while sound and Alfvén speeds may approach the speed of light, results about synchrotron emission and polarization properties are expected to be similar for classic MHD turbulence runs too, readily applicable to the case of SNRs.
We hope that the present work may help to connect ongoing (and future) X-ray polarimetric observations of any kind of high-energy astrophysical source to properties of magnetic turbulence.

\begin{acknowledgements}
The authors thank Rino Bandiera for unvaluable discussions, started in June 2024 during the workshop at Villa Il Gioiello at Arcetri to celebrate his retirement, and the anonymous referee for the positive and constructive review. We acknowledge support from PRIN-MUR project 2022TJW4EJ and from the ICSC — Centro Nazionale di Ricerca in High-Performance Computing; Big Data and Quantum Computing, funded by European Union - NextGenerationEU. All simulations were performed on the Leonardo supercomputer at CINECA (Bologna, Italy). LDZ acknowledges CINECA for the availability of HPC resources through a CINECA–INAF agreement (allocation INA24\_C3B05) and through a CINECA–INFN agreement (allocation INF24\_teongrav). 
\end{acknowledgements}

\bibliographystyle{aa.bst}

\appendix

\section{Correction for anisotropy of magnetic fluctuations to the analytical estimate of the PD}
\label{app:nicco-true-model}

We show here how the results by \citet{Bandiera2016} can be extended to account for anisotropic turbulence, even in the presence of a background magnetic field (at least in the limit of a small level of such anisotropy). To begin with, one needs to remember that only magnetic fluctuations perpendicular to the mean background magnetic field $\boldsymbol{B}_0$ contribute to the depolarization of the signal, while both magnetic fluctuations parallel and perpendicular to it contribute to the total intensity. We recall that in the case considered for the PD analytical prediction we have a mean field fully contained in the plane of the sky, hence $\bar{B} \equiv B_0$ (anyway here we prefer to use $B_0$ for sake of clarity), so we expect that both $\delta B_\parallel$ and just one component $\delta B_\perp$ contribute to the total intensity, while $\delta B_\perp$ alone enters in the polarized intensity (coincident with $\delta B/\sqrt{3}$ in the isotropic case).

Let us then assume that magnetic fluctuations are not isotropic, in the sense that the PDF width of $\delta B_\parallel$ is less than that of $\delta B_\perp$ (as shown in Fig.~\ref{fig:anisotropy}, in selected cases) and parametrize this fact by introducing an effective anisotropy coefficient
\be
a_{\rm eff} = \frac{\delta B_\perp^2 -\delta B_\parallel^2}{\delta B^2},
\label{eq:aeff-def}
\ee
which only takes positive values, as invariably observed in our simulations, but it is  not necessarily a constant for all cases (in principle it may depends on the level of initial fluctuations, the plasma beta, the level of intermittency of turbulence, and so on). In the above expression $\delta B^2 = 2\delta B_\perp^2 +\delta B_\parallel^2$, assuming that the variance in the two perpendicular directions is the same (as we also find in simulations, approximately). Now, by following similar arguments already present in \citet{Bandiera2016} (but beware of the different sign in the definition of the anisotropy coefficient), an effective way to predict the PD by using the same analytical formulas of Eq.~(\ref{eq:BP16}), through the definition of $\xi$, or of Eq.~(\ref{eq:DZ25}) is to replace, at the same time, the background field intensity $B_0$ with a new (lower) $B_{\rm eff}$, and the isotropic $\delta B$ with a new (higher) $\delta B_{\rm eff}$, respectively defined as
\begin{eqnarray}
B_{\rm eff}^2 & = & B_0^2 + \delta B_\parallel^2 - \delta B_\perp^2 = B_0^2 -a_{\rm eff}\, \delta B^2, \\
\delta B_{\rm eff}^2  & = & \delta B^2 + \delta B_\perp^2 - \delta B_\parallel^2 = (1+a_{\rm eff})\, \delta B^2.
\end{eqnarray}
Hence, one can replace the ratio of these two quantities, appearing in the argument of the analytical expressions, using
\be
\frac{B_0^2}{\delta B^2}\rightarrow \frac{B_{\rm eff}^2}{\delta B_{\rm eff}^2} = \frac{1}{1 + a_{\rm eff}}\left(\frac{B_0^2}{\delta B^2}- a_{\rm eff}\right),
\ee
which reduces to the isotropic case in the limit $a_{\rm eff}\rightarrow 0$, by construction. 

In general however, the anisotropy in Eq.~(\ref{eq:aeff-def}) is itself a function of the ratio $\delta B/B_0$, being lower for stronger (normalized) perturbations. We find that a reasonable ansatz, capable of fitting better the results of Fig.~\ref{fig:Rino-plot}, is to choose  
\be
a_{\rm eff} \simeq \frac{a_0}{1 + (\delta B/B_0)^2},
\label{eq:aeff-func}
\ee
where the parameter $a_0$ is the limit for $\delta B \ll B_0$. Notice that when $\delta B/B_0 < 1$ (and $a_0 < 1$) it is easy to show that the combination of the above expressions leads to 
\be
\frac{B_0^2}{\delta B^2}\rightarrow \frac{B_{\rm eff}^2}{\delta B_{\rm eff}^2} \simeq \frac{1}{1 + a_0}\frac{B_0^2}{\delta B^2} \simeq (1 - a_0)\frac{B_0^2}{\delta B^2},
\ee
so that one is basically allowed to enhance by a constant factor $1/(1-a_0)$ the ratio $(\delta B/B_0)^2$ appearing in the analytical expressions. The same is also true in the opposite limit $\delta B/B_0 > 1$ (and again $a_0 < 1$), so this is basically a universal behavior. When applied to the simpler formula DZ25 for the PD estimate, this correction is equivalent to substitute
\be
A \to A_{\rm eff} = \frac{A}{1 - a_0},
\ee
leading to slightly higher values for $A$, with respect to the fit valid for isotropic fluctuations, as anticipated when discussing the dashed line in Fig.~\ref{fig:Rino-plot}.

\begin{figure}[t]
    \centering
    \includegraphics[width=90mm]{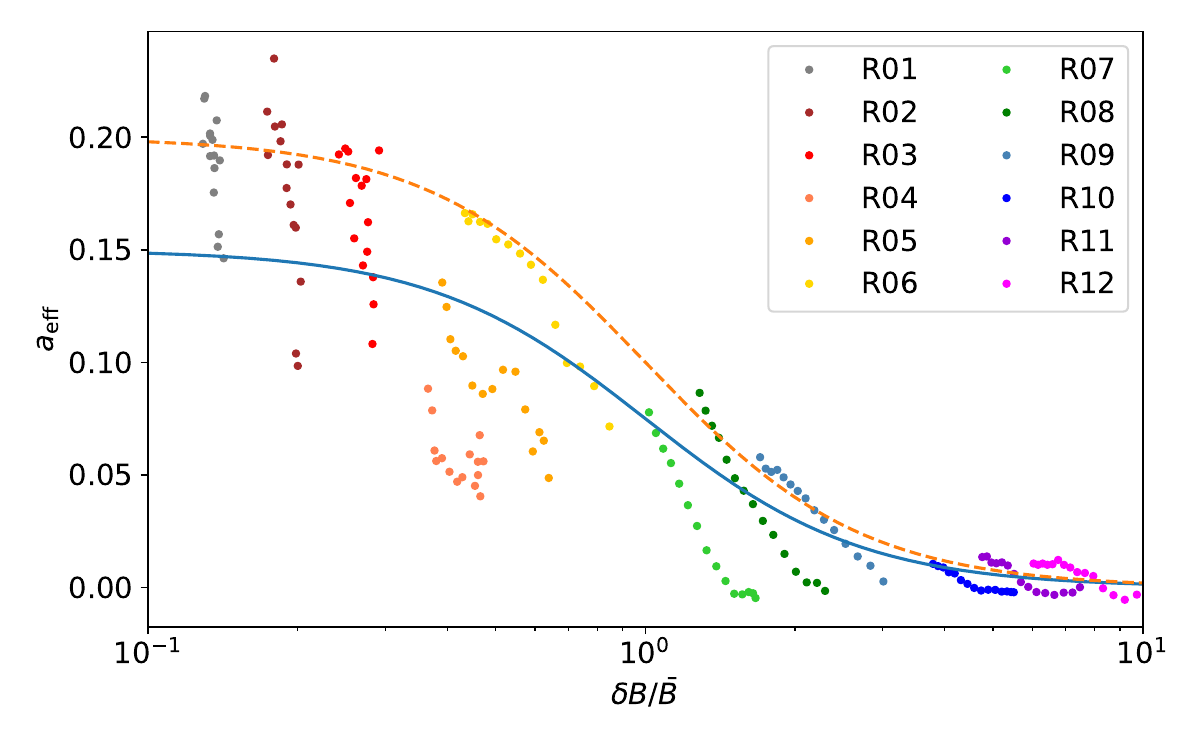}
    \caption{Comparison of the level of anisotropy in our simulations computed as in Eq.~(\ref{eq:aeff-def}), with the analytical function in Eq.~(\ref{eq:aeff-func}).
    We use a solid blue line for the value of maximum anisotropy $a_0 = 0.15$ (our best fit value), and a dashed orange line for $a_0 = 0.20$ (providing a sort of safety upper bound for all simulation outputs). }
    \label{fig:Nicco-plot}
\end{figure}

In Fig.~\ref{fig:Nicco-plot} we compare our analytical expression in Eq.~(\ref{eq:aeff-func}) against the computed $a_{\rm eff}$, using  Eq.~(\ref{eq:aeff-def}), on top of results from all our numerical simulations. Each dot corresponds to an output time, the same choices as in Fig.~\ref{fig:analytical}, from slightly before the peak of turbulence to the end of the run. As in the mentioned figure, as time increases, normalized fluctuations decrease, hence time increases toward the left. Notice how the above choice for the function $a_{\rm eff}$ (roughly) reproduces the general trend observed within our simulations, in spite of the spread of the results being quite large, even for the different outputs of a single run (suggesting that the development of the turbulent cascade and the increasing level of intermittency may play an additional role). The value of $a_0=0.15$ (corresponding to the function with the blue solid line) leads to $A_{\rm eff} \simeq 0.85$ (modifying the original value $A=0.72$), and corresponds to the best fit for anisotropic turbulence of Fig.~\ref{fig:Rino-plot}. Raising the coefficient $a_0$ from 0.15 to 0.20 (the yellow dashed curve) provides an upper bound to our numerical results, but does not change much the trend in terms of the PD, so we prefer to adopt $a_0 = 0.15$.

\end{document}